# Criticality-related fundamental bases for new generations of gas-liquid, liquid-liquid, and liquid (LE) extraction technologies.

Sylwester J Rzoska and Aleksandra Drozd-Rzoska

Institute of High Pressure Physics Polish Academy of Sciences,

ul. Sokołowska 29/37, 01-142 Warsaw, Poland

**Abstract**

The report presents experimental results, which can be considered as the reference for innovative generations of Supercritical Fluids (SCF), Liquid-Liquid (LL), and Liquid (L) extraction technologies. They are related to implementations of *Critical Phenomena Physics*, for such applications not considered so far. For the gas-liquid critical point, the shift $SuperCritical\ Fluids\ (SCF)\ \Rightarrow SubCritical\ Fluids$, due to the additional exogenic impact of ultrasounds is indicated. For LL technology, the possibility of increasing process effectiveness when operating near the critical consolute under pressure is indicated.

Finally, the discovery of long-range precritical-type changes of dielectric constant in linseed oil, standing even 50K above the melting temperature, is presented. It suggests that extraction processes linking 'SCF' and 'L' technologies features, and exploring the natural carrier, such as linseed oil, is possible. The report recalls the fundamental base for extraction processes via Kirkwood and Noyes-Withney relations, and presents their implementations to 'critical conditions', including pressure.

## 1. Introduction

The name ‘supercritical fluids’ (SCF) can be linked to any liquid beyond (above) the critical point, in the homogeneous domain. For the SCF extraction technology, it is associated with the vicinity of the gas-liquid critical point, i.e., the domain above the critical pressure ($P_C$) and temperature ($T_C$). SCF often explores carbon dioxide as the ‘criticality carrier’, with a small addition of co-solvents to focus on the desired extraction target [1-3].

However, the most commonly used extraction technology is related to the dissolution of a selected component in a liquid solvent. Such liquid (L) extraction technology is used in numerous industrial applications and everyday life: in the kitchen when preparing dishes and even during the digestive process in our body. Due to larger densities of liquids, in comparison to the gas state, liquid solvents are characterized by higher solvation power than SCF technology: more solvent molecules are exposed to solute molecules. However, the ‘solvent power’ can be in the given case so high that it reduces the extraction process selectivity, which can be a significant disadvantage [4]. The strong cooling can offer some control of this factor, but it also decreases the process effectiveness and raises costs. The so-called ‘botanical extraction’, based on methanol cooled down to −70 $^{o}$C, can be recalled as an example [5]. The search for natural and non-toxic carrier liquids constitutes the essential development direction for the ‘L’ extraction technology [4, 5].

Liquid-liquid (LL) extraction technologies, also known as partitioning, are based on different solubilities of one component of the processed system in two different immiscible but coexisting liquid phases, separated by a meniscus [6-9]. Water as the polar component and a non-polar organic solvent are most often used in applications. During processing, a net mass transfer of species from the first liquid to the second one occurs. The solvent, enriched in solute(s), is called ‘extract’. The ‘fed solution’, depleted in solute(s), is called the ‘raffinate’ [5-7]. Regarding future developments, the application of the limited miscibility phenomenon is indicated [6-9].

The industrial significance of extraction technologies is enormous, from perfume & cosmetics, pharmaceutical, food, biofuel, and fine chemicals industries to hydrometallurgy and biotechnology [1-9]. Notwithstanding, each extraction technology has significant advantages and disadvantages. The SCF gas-liquid critical point extraction technology is the most selective and fine-tuning method. Hence it is often indicated as the most promising 'green' approach. However, the inherent usage of high-pressure and gas-based technical solutions, matched with the high-level control of parameters, can restrict applications, particularly for large amounts of processed materials. Liquid-based extraction technologies (L, LL) are formally more effective but notably less selective. They often require strong cooling or heating, which raises costs. The significant problem constitutes post-process pollution.

This report presents evidence indicating fundamental bases for new generations of extraction technologies. They are related to the sub-criticality extending the classic SCG approach and the critical-type developments of 'traditional' liquid (L) and liquid-liquid (LL) extraction technologies solutions.

## 2. Critical phenomena, dielectric properties, and their link to basic characterizations of the extraction process

The *Physics of Critical Phenomena* was the grand universalistic success of the 20$^{th}$ century physics, chemical physics, and material sciences [10, 11]. It enabled the explanation of common patterns appearing for qualitatively different systems, at the microscopic level, on approaching the critical temperature ($T_C$, continuous phase transition). Such behavior is governed by universal critical exponents, which values depend solely on the space ($d$) and order parameter ($n$) dimensionalities. The pretransitional/precritical changes of physical properties are shaped by multi-molecular or more general multi-element critical fluctuations, which lifetime $(\tau_{fl.})$, and size (correlation length, $\xi(T)$) are described as follows [10, 11]:

$$\xi(T) = \xi_0^P (T - T_C)^{-\nu} \ , \qquad \tau_{fl.}(T) = \tau_0^P (T - T_C)^{-\phi=-z\nu} \ \ P = const \qquad (1)$$

where $T_C$ and $P_C$ are for continuous ('critical') phase transition temperature and pressure. Near critical systems can be organized into universality classes $(d, n)$, for which pretransitional effects are described by power-type dependences with the same values of critical exponents. For this report, significant is $(d = 3, n = 1)$ universality class to which belongs: (i) the surrounding of the gas-liquid critical point, (ii) the critical consolute point in binary mixtures of limited miscibility, and (iii) the critical point in 3-D Ising model, or paramagnetic – ferromagnetic Curie temperature. For this universality class, the correlation length critical exponent in Eq. (2) $\nu \approx 0.625$ and the dynamic exponent has the value linked to the conserved order parameter $z = 3$.This yields for the collective relaxation tome the exponent $\phi \approx 1.88$. As for other critical exponents $\alpha \approx 0125$, $\beta \approx 0.325$, $\gamma \approx 1.2$ for the specific heat (heat capacity), order parameter, and the order parameter related susceptibility, respectively. Only two critical exponents are independent: they are linked via scaling equations. Generally, the mentioned single power term precritical description obeys only very close to the critical temperature: in practice for $\Delta T_C < 1K$. On moving away, the additional power terms associated with so-called corrections-to-scaling exponents are required. The same values of critical exponents are obtained when approaching the singular, critical temperature in sub- and super-critical domains, i.e., for $T < T_C$ and $T > T_C$, respectively. The mentioned characterizations are for so-called non-classical critical phenomena [10, 11].

When increasing the space dimensionality, for $d \geq 4$, values of exponents are strictly defined, namely: $\nu = 1/2$, $\gamma = 1$, $z = 2$ (the value for the non-conserved order parameter), $\gamma = 1$, and $\beta = 1/2$. The unique situation occurs for the specific heat-related exponent: $\alpha = 0$ for $T > T_C$ and: $\alpha = 1/2$ for $T < T_C$ in the mean-field type systems. The 'border dimensionality' is reduced to $d \geq 3$ for the tricritical point (TCP) case, associated with the meeting of three phases at the TCP singularity. In the given case following values of exponents appear: $\nu = 1/2$, $\gamma = 1$, $\beta = 1/4$ and $\alpha = 1/2$ for the low- and high- temperature sides of TCP. It is the case of 'classical' critical phenomena: mean-field, or TCP type. Notable that supplemental correction-to-scaling terms in the description of precritical phenomena are absent. A single power term can describe pretransitional behavior up to a few tens of Kelvins away from the singular temperature [11].

The critical phenomena isomorphism postulate suggests the same 'universal' precritical behavior on approaching the critical point using any field-type ('intensive) thermodynamic variable. As an example can serve temperature or pressure [11]. The additional variable field-type ('intensive') or density type ('extensive') can shift up or down the critical temperature in the given system. Notable that the density-type disturbations influence the values of critical exponents [11]. For the simplest case of a solvent 'dissoluting' the system without significant interaction with its components, the so-called Fisher renormalization takes place [12, 13]:

$$\gamma \Rightarrow \frac{\gamma}{1-\alpha} \quad , \quad \beta \Rightarrow \frac{\beta}{1-\alpha} \quad , \quad \alpha \Rightarrow \frac{-\alpha}{1-\alpha} \qquad (2)$$

For this report, essential is the precritical behavior of dielectric constant. The relation describing its behavior in the supercritical, homogeneous domain above the gas-liquid critical point or the critical consolute temperature binary mixtures of limited miscibility was derived by Sengers et al. [14] via the analysis of the internal energy under a slightly disturbing electric field:

$$\varepsilon(T) = a(T - T_C) + A(T - T_C)^{1-\alpha}[1 + c(T - T_C)^{\Delta_1} + \cdots] \qquad (3)$$

where $T > T_C$ the exponent $\alpha \approx 0.12$ and the first correction-to-scaling exponent $\Delta_1 \approx 0.5$ Similar behavior was found for the isotropic liquid phase (I) of rod-like liquid crystalline (LC) mesophasic behavior on approaching the nematic (N), chiral nematic (N*), or smectic (SmA, SmE) phases [15-20]:

$$\varepsilon(T) = a(T - T^*) + A(T - T^*)^{1-\alpha} \tag{4}$$

where $T > T_{LC}$, the latter is for the weakly discontinuous 'melting' phase transitions, the singular temperature of a hypothetical continuous phase transition $T^* = T_{LC} - \Delta T^*$, and the exponent $\alpha = 1/2$ .

It is the case of a weakly discontinuous phase transition in LC systems, in which the discontinuity metric $\Delta T^*$ can change from $1 - 2K$ for the isotropic liquid – nematic (I-N) transition to $\sim 30K$ for the isotropic liquid – Smectic E (I-SmE) transition [15-20]. The description of the 'supercritical' behavior via Eq. (4) in LC materials was introduced heuristically, using the similarity to the pretransitional behavior in critical mixtures (Eq. (3)) [15-20]. Only recently, the model deriving Eq. (4) and covering the isotropic liquid phase of liquid crystals, and also plastic crystals has been proposed [21].

Amongst physical properties which are significant for extraction technologies worth recalling is DC electric conductivity. It is the dynamic dielectric property for which the Arrhenius or Super-Arrhenius (SA) behavior is generally expected [22]:

$$\sigma(T) = \sigma_\infty exp\left(-\frac{E_a(T)}{RT}\right) \tag{5}$$

where $R$ stands for the gas constant and $E_a$ denotes the activation energy. It takes the temperature-dependent, apparent $E_a(T)$ form for the SA dynamics and the value $E_a = const$ in the given temperature domain for the basic Arrhenius behavior.

Notwithstanding studies in the homogeneous binary critical mixtures of limited miscibility revealed the temperature evolution parallel to Eq. (4) [23-25]:

$$\sigma(T) = a_\sigma(T - T_C) + A_\sigma(T - T_C)^{1-\alpha}[1 + d_\sigma(T - T_C)^{\Delta_1} + \cdots] \tag{6}$$

Both dielectric constant and DC electric conductivity can be determined from broadband dielectric spectroscopy (BDS) scans [22]. They constitute leading parameters in relations significant for extraction processes. One can recall the classic Kirkwood relation [26, 27]:

$$k, s = p_\infty exp\left[\frac{A\Delta p}{RT}\left(\frac{1}{\varepsilon} - 1\right)\right] \tag{7}$$

where $k$ and $s$ stand for the chemical reaction rate and the solubility; $\Delta_P$ is for the difference in polarity between the reactant and product; $p_\infty$ is the prefactor related to the given property, $A$ is the system-dependent constant.

For the extraction process, the solute dissolution rate in time $dm/dt$ is also essential. It is associated with the diffusion rate, as shown by Noyes-Whitney equation [28, 29]:

$$\frac{dm}{dt} = S\frac{D}{d}(C_S - C_B) \tag{8}$$

where $m$ is the mass of dissolve material, $t$ – the processing time, $S$ is the surface area of the solute particle, $D$ diffusion coefficient, $d$- is the thickness of the concentration gradient layer, $C_S$ and $C_B$ particles surface and bulk concentrations ($mol/L$).

One can relate the translational diffusion in Eq. (8) to the DC electric conductivity using the Debye-Stokes-Einstein relation [30]:

$$D_{trans.} = \frac{k_B T}{nq^3}\sigma = KT\sigma \tag{9}$$

where $n$ is for the number of electric carriers/charges ($q$) and $K = k_B/nq^3$

One can  expect that 'critical systems' the evolution of $dm/dt$ depends on the temperature distance from the critical point. Linking Eqs. (6-9) one obtains:

$$\frac{dm}{dt} = D\frac{S}{d}(C_S - C_B)\, KT\sigma(T) = \; D\,\frac{S}{d}\Delta_C\, KT\sigma(T) \;\Rightarrow \tag{10a}$$

$$\Rightarrow \;\; \frac{dm}{dt} = CT\left(a_\sigma(T - T_C) + A_\sigma(T - T_C)^{1-\alpha} + \cdots\right) \tag{10b}$$

where $C = (S\Delta_C/d)(k_B/nq^3) = const.$

For the 'critical approximation' of the Kirkwood relation (Eq. (9)), one can derive:

$$k(T), s(T) \approx \frac{Ap_\infty \Delta_P}{RT}\left(\frac{1}{\varepsilon(T)} - 1\right) + p_\infty = \frac{K}{T}\left(\frac{1}{\varepsilon_C + a(T-T_C) + A(T-T_C)^{1-\alpha} + ..} - 1\right) + p_\infty \tag{11}$$

where the constant parameter $K = Ap_\infty \Delta_P/R$

In the above relation, it is assumed $P \approx const$, which can be valid even for SCF technology, implemented in a well-defined and limited region above the critical temperature and pressure. Hence, for $k(T), s(T)$ behavior near the critical point, the precritical behavior 'opposite' to the one observed for $\varepsilon(T \rightarrow T_C, T^*)$ appears. For instance, in critical mixtures or LC systems in the homogeneous (i.e., Isotropic) liquid phase, one observes the anomalous decrease of $\varepsilon(T)$ for $T_C \leftarrow T$, and then the precritical increase for $k(T), s(T)$ may be expected.

## 3. Experimental

This report is based on new and earlier experimental results of the authors. The latter have been up-to-date, supplemented experimentally, and reanalyzed, focusing on the report's target. Experimental studies were carried out via the broadband dielectric spectroscopy (BDS) impedance analyzer Novocontrol. Tested samples were BDS scanned to obtain complex dielectric permittivity $\varepsilon^* = \varepsilon' - i\varepsilon''$ spectra: examples are shown in Figure 1, for the linseed oil in the liquid and solid phases. The plot also presents characteristic features related to properties significant for the discussion. The real component of dielectric permittivity was determined as $\varepsilon'(f) = C(f)/C_0$, where $C$ is the capacitance for the measurement capacitor filled with the tested dielectric and $C_0$ is for the empty capacitor. The imaginary part was calculated as $\varepsilon''(f) = 1/2\pi f R(f) C_0$ , where $R(f)$ stands for resistivity [22]. The horizontal part of $\varepsilon'(f)$ in Fig. 1 is for the static domain, related to dielectric constant. It is composed of $\varepsilon_\infty$ (linked to atomic and molecular polarizabilities) and $\varepsilon_S$ associated with dipolar polarizability, i.e., $\varepsilon = \varepsilon_S + \varepsilon_\infty$. The dipolar component diminishes on increasing the frequency: in the given case for $f > 300kHz$ (Fig. 1a). For low molecular weight liquids, this domain can start even for. For the low-frequency part, usually $f < 1kHz$, the strong impact of polarizability associated with the translation of ionic species emerges, and the estimation the DC electric conductivity is possible: $\sigma = \omega\varepsilon''(f) = 2\pi f \varepsilon''(f)$ [22]. For the log-log presentation of the spectrum, as in Fig. 1b, DC electric conductivity is related to the 'thick, grey' line described as follows:

$$log\varepsilon''(f) = lg\sigma - d \times log\omega \qquad (12)$$

where $\omega = 2\pi f$, and $d = 1$ for DC electric conductivity

For the coefficient $d \neq 1$, other mechanisms contribute to the electric conductivity, such as the Maxwell-Wagner polarization of electrodes [22]. Such effects are beyond the given report.

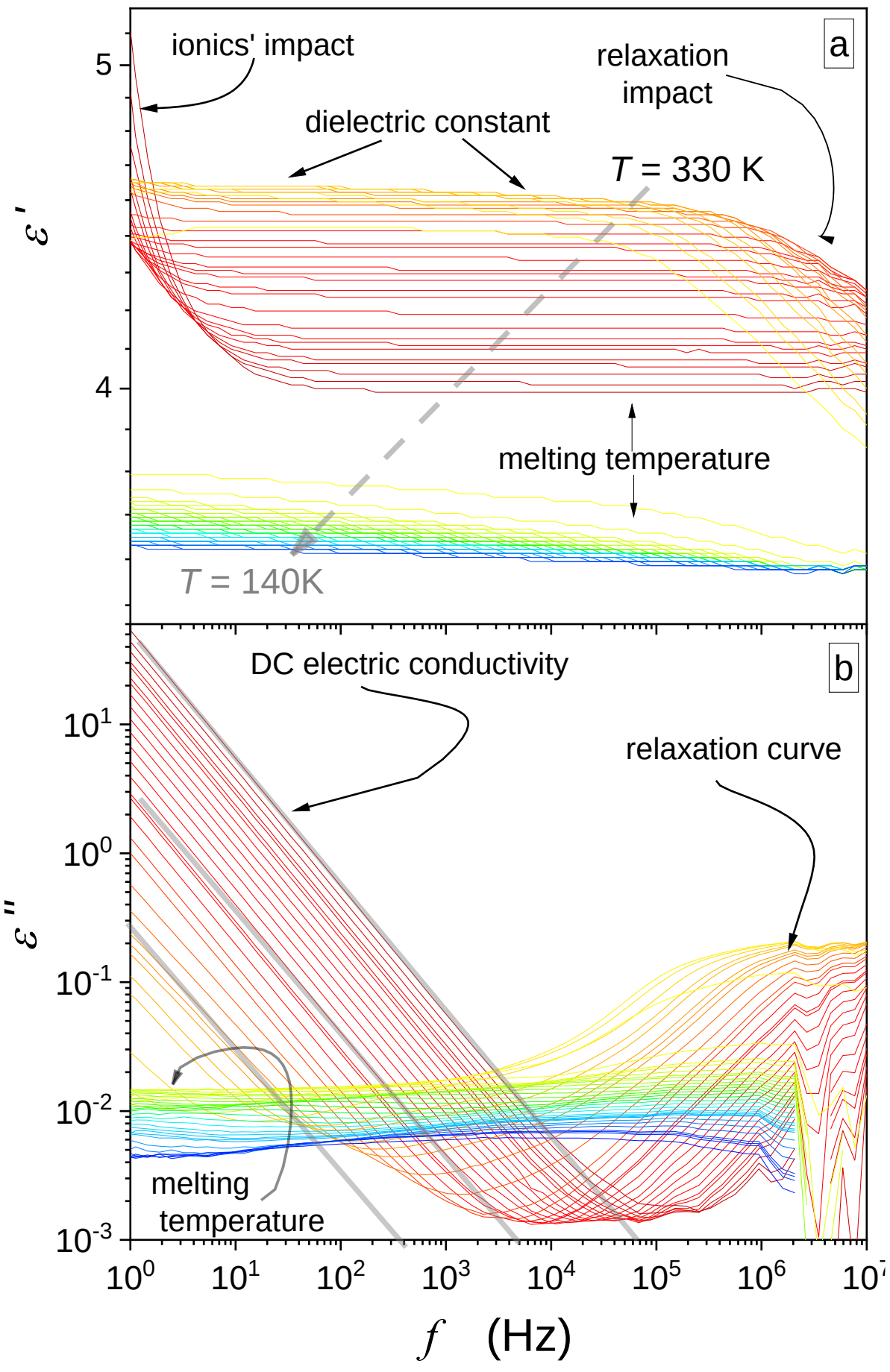

**Fig. 1** Selected BDS spectra expressed via frequency changes in the real and imaginary parts of dielectric permittivity $\varepsilon^* = \varepsilon' - i\varepsilon''$, in linseed oil, for the range of temperatures covering liquid and solid phases. Domains of spectra related to significant dielectric properties are indicated. Grey, thick lines show the 'preserved' DC electric conductivity condition, with slight distortions appearing only in the immediate vicinity of the melting temperature.

Results presented in Fig. 1, are for the 'hot', new scan for linseed oil, what is the reason for some disturbations for $f > 1MHz$, although the meaning of the relaxation domain remains visible. However, the given report focuses on the evolution dielectric constant in the static domain, generally occurring in the frequency range between $\sim 1kHz$ and $\sim 300kHz$. Fig. 1 illustrates key features of the complex dielectric permittivity-related BDS scans, including possible experimental problems. The latter situation (not important for the given report) can be solved by the cumulative analysis and/or better adjustment of the sample impedance to the scanning window of the analyzer, as a rule narrowing near the analyzer frequency borders. Note that this report is related to changes in dielectric constant, which is the physical property significant for extraction processes. Similar pretransitional anomalies can also be noted for other properties, such as refractive index ($n$) or density ($d$), but they are not significant for the target of this report. Regarding the basics of the *Physics of Critical Phenomena*, it is notable that pretransitional anomalies for dielectric constant often manifest 'stronger', than for $n(T)$ or $d(T)$ changes, significantly supporting the reliable analysis.

Studies presented below are for Diethyl Ether (DEE) with the gas-liquid critical point [31, 32], nitrobenzene (Nb) – dodecance (Dd) mixture of limited miscibility with the critical concentration [25], and linseed oil [33]. DEE, Nb, and Dd were purchased from Fluka, with the highest available purity. Nb and Dd were additional twice distilled under reduced pressure. From widely cultivated in Poland, Golden flax, *Linum usitatissimum*, seeds were used to prepare the linseed oil via their cold-pressing in the laboratory. Seeds were purchased from Herbapol S.A., Lublin, Poland: the leading polish company specializing in high-quality pro-health seeds & herbs products. The triple filtering of the oil, using coffee filters was carried out.

## 4. Results and Discussion

Dielectric properties are important for extraction processes due to their direct link to basic process characterizations, as indicated above. Near the critical point, physical properties anomalously change, and they are governed by precritical/pretransitional phenomena. Notwithstanding the experimental evidence of supercritical anomalies of such dielectric properties for the gas-liquid critical point., essential for SCF technology remains limited. One can recall studies in CO [34], $SF_6$ [35], and diethyl ether (DEE) [31, 32]. Results for the latter, carried out by the authors of this report, are presented in Fig. 2. The unique isochoric path of scanning is notable. It facilitates the high-level control of parameters describing the distance from the critical point. Distanced by 30 mm, two measurement capacitors were placed in the pressure chamber. To the remaining volume within the pressure chamber ($V_{free} = 21.4\ cm^3$) such amount ($V, m$) of DEE was added that $V_{free}/m = \rho_C = 0.263\ g/cm^3$, i.e., it was equal to the critical density. At room temperature, the density of DEE is equal to $\rho = 0.713\ g/cm^3$, [32] hence ca. $1/3$ of $V_{free}$ was filled with the tested liquid, covering one of the measurement capacitors. Subsequently, the chamber was closed and slowly heated. On raising the temperature, thermodynamic characterizations of the system $(T, P)$ changed along the critical isochore, reaching the critical pressure at $T = T_C$. Subsequently, for $T > T_C$, both capacitors are in the homogeneous phase. During the experiment, the temperature was shifted up to values well above $T_C$. For the isochoric path, temperature changes are paralleled by pressure ones. Notable is the precritical increase of dielectric constant for $T_C \leftarrow T$, fairly well portrayed by Eq. (3) with the following parameters: $\varepsilon_C = 1.45$, $A = -0.155$, $a = 0.012$, $c = 0.04$, $\alpha = 0.12$.

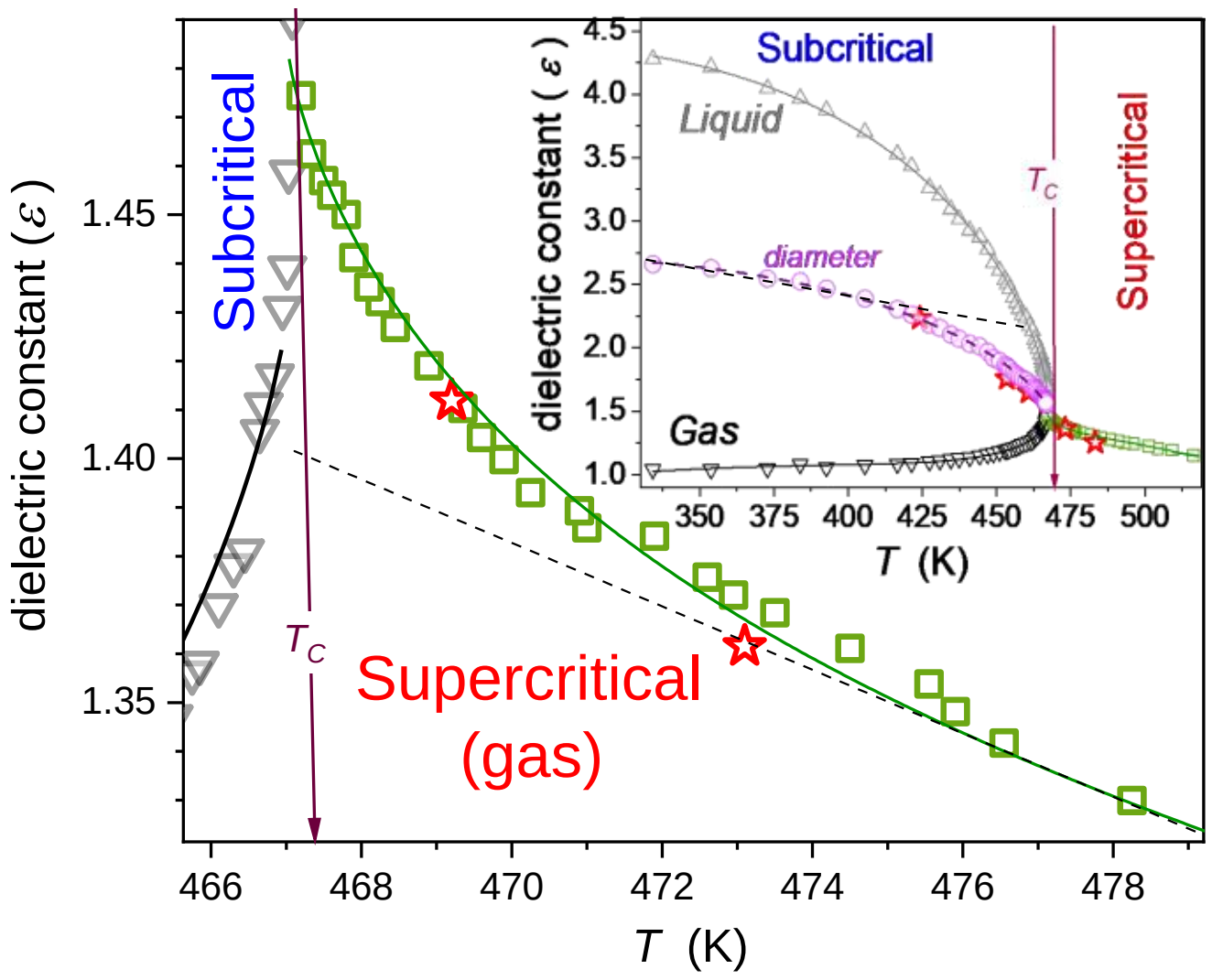

**Fig. 2** Temperature dependence for isochoric approaching the gas-liquid critical point in diethyl ether [25]. The inset shows full range results, including also the gas-liquid

equilibrium domain. Violet points and the dashed curve show the parameterized evolution in the two-phase domain. Stars show results of direct measurements of the system under the additional impact of the ultrasound, supporting the 'homogenization'. The critical temperature $T_C = 467K$ is linked to the critical pressure $P_C = 3.6MPa$. The dashed line in the inset presents the 'classic' Cailletet-Mathias (CM) law of rectilinear diameter behavior: see comments below and ref. [36-38]. The strengths of pretransitional anomalies for the diameter ($T < T_C$)and in the homogenous (isotropic) phase ($T > T_C$) can be estimated as $\Delta_d = 40\%$ and $\Delta_{iso.} = 5\%$ (see comments below). The dashed lines are for the CM law of rectilinear diameter, and approximately linear changes (within the limit of the experimental error) of dielectric constant in the homogeneous region.

The pretransitional anomaly in the homogeneous supercritical domain ($T > T_C$) is smoothly linked to the anomaly of the diameter of the coexistence curve (two-phase domain) for $T < T_C$. Its pretransitional anomaly is well portrayed by the relation [39]:

$$d(T) = \frac{\varepsilon_{liquid}+\varepsilon_{gas}}{2} = \varepsilon_C + B_d(T_C - T)^{2\beta} + A_d(T_C - T)^{1-\alpha} + a_d(T_C - T) + \dots \quad (13)$$

where $T < T_C$ and $B_d = -0.056$, $A_d = 0.15$, $a_d = -0.064$; for critical exponents $\alpha = 0.12$ , $\beta = 0.325$

When comparing the above anomalies in the supercritical ($\varepsilon(T)$) and subcritical domains ($d(T)$), the latter is essentially stronger and extent, well remote from $T_C$. It is particularly well visible in Fig. 3, presenting the derivative plot based on data from Fig. 2. It is worth stressing that Fig. 3 recalls the link between the derivative of dielectric constant and the heat capacity anomaly, both for the homogeneous region and the two-phase domain, which can be concluded from refs. [37, 40]. For $T < T_C$, strong pretransitional changes start already at $T_1 \approx T_C - 60K$, whereas for $T > T_C$ the onset is related to $T_2 \approx T_C + 10K$. It can suggest that using the diameter of the coexistence curve in the subcritical domain as the reference for the Subcritical Fluids extraction process can be more efficient and better controlled than the classic treatment for the gas-liquid Supercritical domain. However, the diameter of the coexistence curve is a ‘functional, calculated’ property, which seems to be experimentally directly non-accessible.

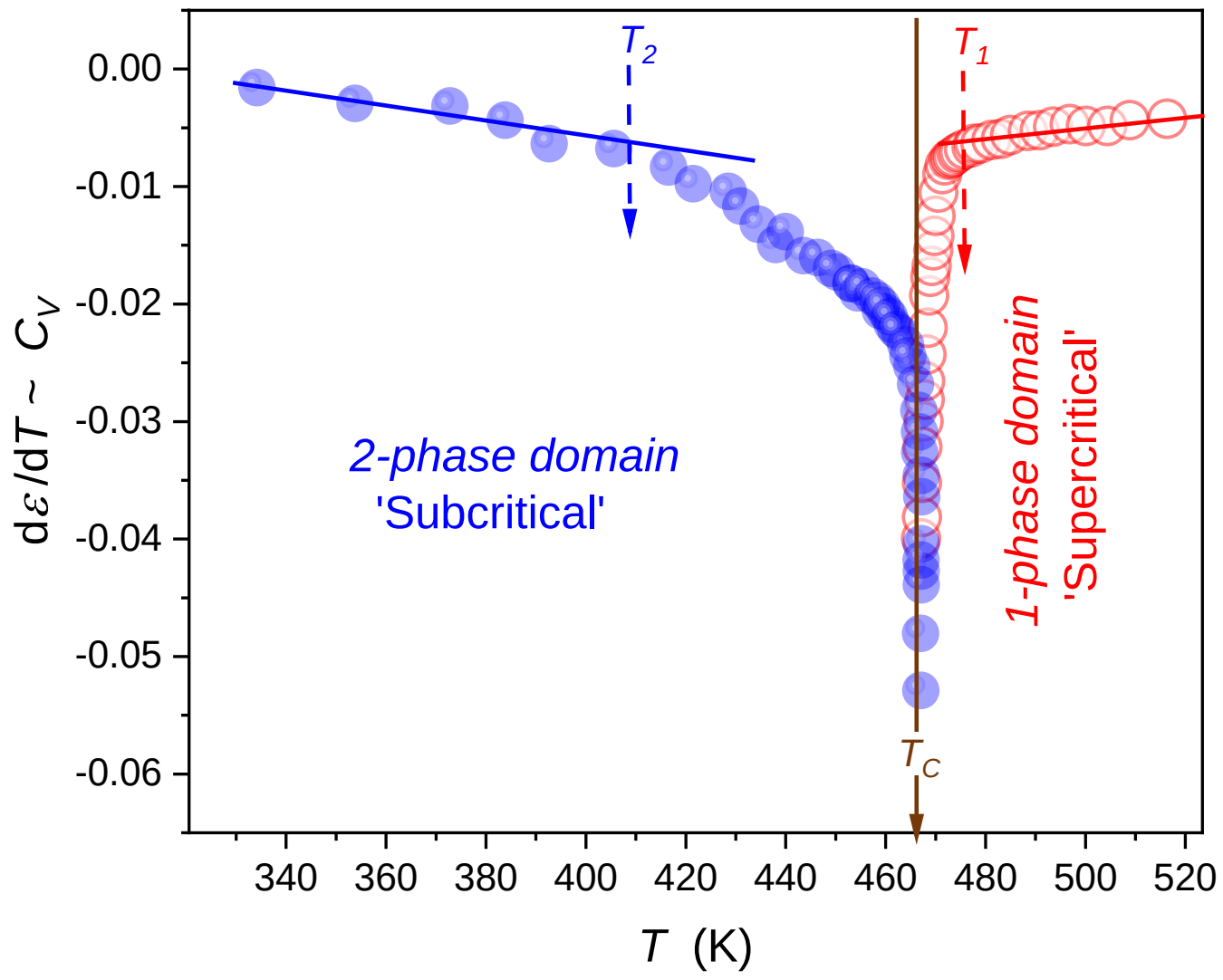

**Fig. 3** The behavior of the derivative of dielectric constant, proportional to heat capacity changes, in the supercritical and subcritical domains. The latter is presented for the diameter of the coexistence curves. Dashed arrows, noted as $T_i$ indicate onsets of the region of strong changes in approaching the critical temperature. As for their ranges $\Delta T^{(1)} = |T^1 - T_C| \approx 10K$ in supercritical domain and $\Delta T^{(2)} = |T^2 - T_C| \approx 50K$ in the subcritical domain.

Notwithstanding, the direct experimental access to the diameter appeared to be possible. Stars in Fig. 2 are for the tested pressure chamber placed in a bath for which the action of the ultrasound ($f$ = 50 kHz) was applied. Under such treatment, for $T < T_C$ dielectric constant changes detected by both measurement capacitors follow the diameter instead of the binodal's gas and liquid-related branches. The action of the ultrasound ‘mixes’ coexisting phases. We stress this issue once more: the measurement points denoted by stars were carried out during the ultrasound homogenization. Without such action, capacitors detected different values of dielectric constant related to the gas- and liquid coexisting phase, respectively.

Consequently, one can propose the Subcritical Fluid (SubCF) technology, associated with the supplemental ultrasound exogenic impact. Following Figs. (2) and (3), the process can be more efficient in the SubCF than in the SCF domain. The SubCF extraction technology can also be convenient to control efficiency because of its colossal temperature range.

Regarding fundamentals of the pretransitional anomaly in the two-phase regions worth recalling is the classic report by Weiner et al. [37], who carried out temperature-related temperature density tests under atmospheric pressure in, below the gas-liquid critical point in $SF_6$. The authors of ref/ [37] noted the violation of the Caillettet- Mathias (CM) law of rectilinear diameter : $d(T < T_C) = \left(\rho_{liq.} + \rho_{gas}\right)/2 = a + bT$ , which omits the existence of the pretransitional anomaly [36, 38] , and it is often used for estimating parameters at the critical point. Basing on data presented in ref. [37], one can estimate the discrepancy between the CM-extrapolated behavior and real value at $T = T_C$: $\Delta_d = 100\% \times \left|d_C - d_C^{CM}\right|/d_C \approx 0.65\%$. A similar analysis for data presented in Fig. 2 yields $\Delta_d = 100\% \times \left|\varepsilon_C - \varepsilon_C^{CM}\right|/\varepsilon_C \approx 40\%$ . The comparison of $\Delta_d$ values can be considered a qualitative metric of the 'strength' of the pretransitional anomaly. Regarding results for DEE, one can also consider the parallel estimation of the pretransitional anomaly 'strength' in the homogeneous (isotropic) phase, for $T > T_C$. Assuming the extrapolation of approximately linear changes $\varepsilon(T)$ remote from the critical point up to $T = T_C$, one obtains $\Delta_{iso.} = 100\% \times \left|\varepsilon_C - \varepsilon_{linear}^{T \gg T_C}\right|/d\varepsilon_C \approx 5\%$. The qualitative difference between values of $\Delta_d$ for $SF_6$ and DEE can be commented as follows: (**i**) $SF_6$ is the non-polar compound, and DEE is the polar compound, (**ii**) the coexistence curve for DEE is much more asymmetric than in $SF_6$, which can be linked to the fact that Fig. 2 it is related to dielectric constant, which value differs enormously in the liquid and gas phases of DEE, (**iii**) Fig. 2 is for isochoric, also pressure-related, studies.

In Figs. 2 and (4-8) dashed lines show the apparently linear behavior of dielectric constant or DC electric conductivity remote from the critical point, extrapolated to $(T_C, P_C)$. They have been introduced to show values of empirical 'strength' parameters $\Delta_d$ and $\Delta_{iso.}$. The authors stress that it is not associated with linear terms in Eqs. (3, 4. 6, 13) and its development below: these contributions are significant at any distance from the critical point and 'alone' do not describe experimental data, even remote from the critical point.

Liquid-liquid equilibria (LLE) of immiscible or partially miscible liquids constitutes another important fundamental phenomenon used for industrial extraction technologies. It enables the extraction in systems where alternative technologies have problems because of the volatility of treated materials. It is particularly significant in hydrometallurgy, pharmacy, and food industries. It is also crucial for the 'sustainable economy' due to its application in 'urban mining', i.e., the recovery of valuable metals from electronic waste [1, 4, 6]. Generally, liquid-liquid extraction is based on separating the dissolved component from its solvent by transferring it to the second liquid phase. It indicates that the application of liquid mixtures of limited miscibility can be particularly interesting here. In such systems, after heating one obtains the homogenous liquid. The subsequent cooling can return to the two-phase equilibrium, but with the desired component primarily solved in the second, coexisting phase – dominated by the more friendly for other processing liquid. It can be surprising that *Critical Phenomena' Physics* is still relatively weakly implemented to gain the in-deep insight into LL extended technologies, although it can improve the process efficiency and minimize waste 'production'. Below, we present the discussion regarding the behavior of dielectric properties in limited miscibility mixtures, which can be essential for LL extraction processing. It includes the possible innovative role of high-pressure conditions.

Figure 4 presents the evolution of dielectric constant in the homogeneous and two-phase regions of nitrobenzene (Nb) - dodecane (Dd) critical mixture. It is composed of a dipolar component with the large dielectric constant (Nb, $\varepsilon \sim 30$) and the non-polar one (Dd, $\varepsilon \sim 2.5$), a pattern often appearing in LL extraction technology. The pretransitional anomaly in the isotropic liquid phase is well portrayed by the parallel of Eq. (3) [25]:

$$\varepsilon(T) = a(T - T_C) + A(T - T_C)^{1-\alpha}[1 + c(T - T_C)^{\Delta_1} + \cdots] \qquad (14)$$

where $T > T_C$ the exponent $\alpha \approx 0.12$ and the first correction-to-scaling exponent $\Delta_1 \approx 0.5$.

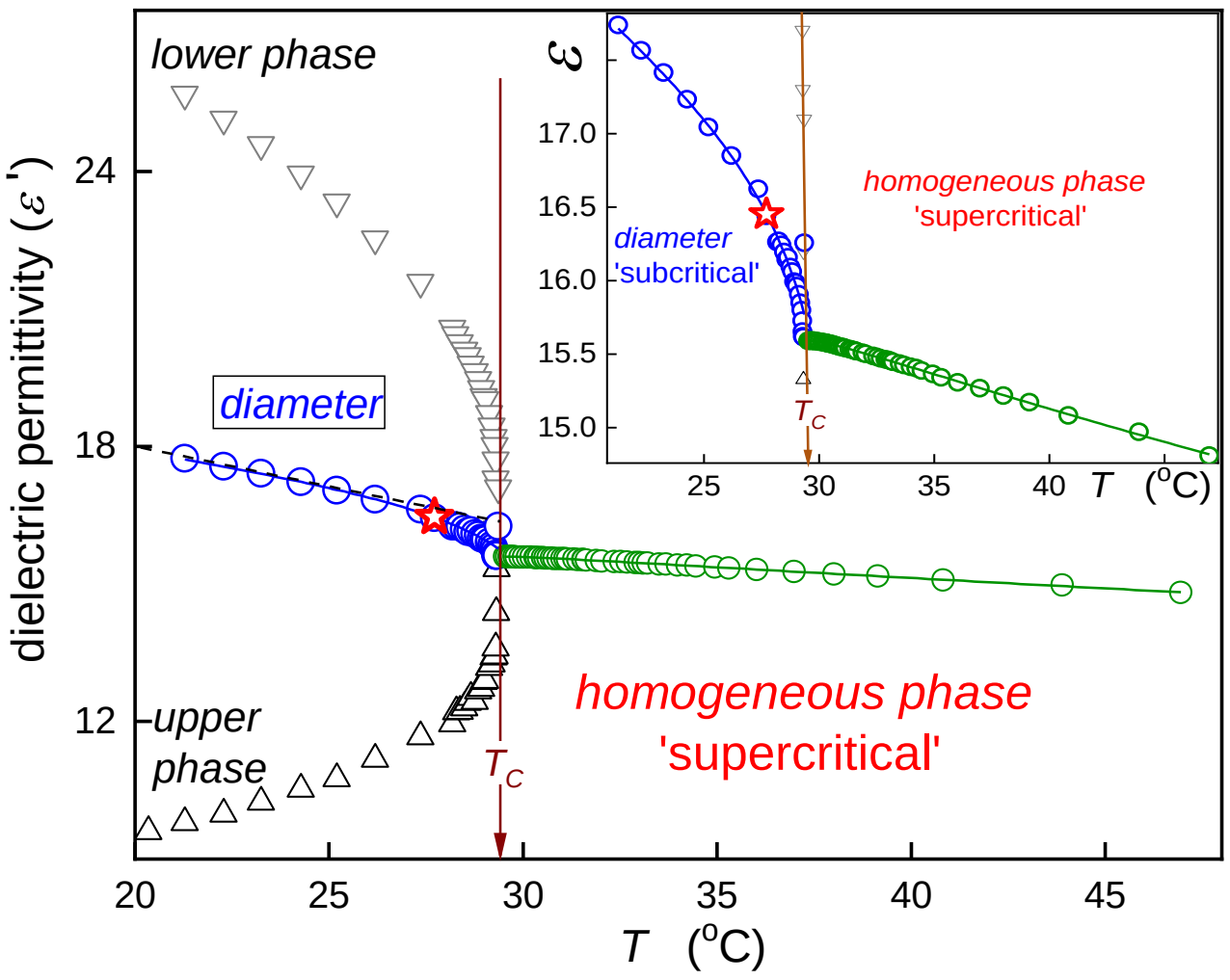

**Fig. 4** Temperature changes, under atmospheric pressure, for dielectric constant below and above the critical consolute temperature in nitrobenzene –dodecane mixture of limited miscibility of critical concentration. The calculated 'diameter' of the coexistence curve is also shown. It is directly available experimentally if ultrasounds additionally influence the system. The inset shows the focused pretransitional behavior in the homogeneous phase, with the portrayal related to Eq. (14). The dashed line is for the CM law of rectilinear diameter, enabling the estimation of the pretransitional anomaly strength $\Delta_d = 5.3\%$; for its parallel in the homogeneous phase $\Delta_{iso.} = 0.3\%$

However, the empirical 'strength' of this anomaly ($\Delta_{iso.}$) is qualitatively lower than for the gas-liquid critical point. For DC electric conductivity, shown in Fig. 5, the pretransitional anomaly is experimentally not detectable. These suggest that the homogeneous phase, for $T > T_C$ and under atmospheric pressure, exhibit only a marginal significance as the 'carrier' for the supercritical extraction – at least if the support of critical phenomena is expected.

Both in Fig. 4 for dielectric constant and Fig. 5 for electric conductivity, the diameters in the two-phase domain show relatively strong pretransitional, 'subcritical', effects. They are described by relations parallel to Eq. (13):

$$d_\varepsilon(T) = \frac{\varepsilon_{liquid}^{upper} + \varepsilon_{liquid}^{lower}}{2} = \varepsilon_C + B_d(T_C - T)^{2\beta} + A_d(T_C - T)^{1-\alpha} + a_d(T_C - T) \qquad (15a)$$

$$d_\sigma(T) = \frac{\varepsilon_{liquid}^{upper} + \varepsilon_{liquid}^{lower}}{2} = \sigma_C + B_d^\sigma (T_C - T)^{2\beta} + A_d^\sigma (T_C - T)^{1-\alpha} + a_d^\sigma (T_C - T) \quad (15b)$$

The diameter is the property supporting the functional analysis of the coexistence curve, formally beyond the direct experimental access. The situation changes when the system is under the additional impact of ultrasound which mixes the upper and the lower coexisting phase, making the system a quasi-homogeneous micro-emulsion. The star in Fig. 4 shows the results of the measurement of dielectric constant for the mixture at $T < T_C$ under the ultrasound exogenic impact. Both measurements capacitors detect the exact value of dielectric constant located at the diameter.

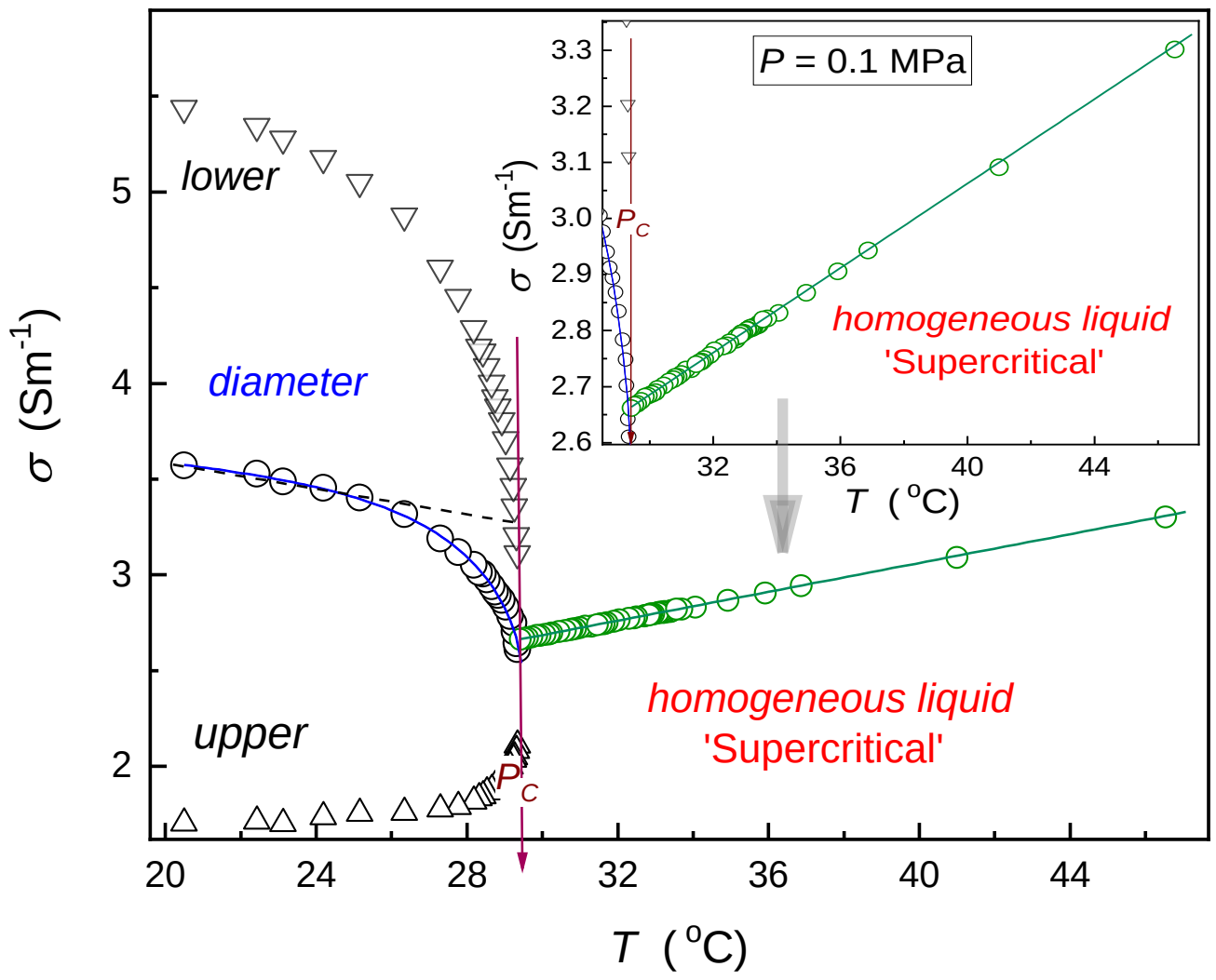

**Fig. 5** Temperature evolution of electric conductivity in nitrobenzene – hexane critical mixture. The behavior of the diameter is also shown. The inset presents the behavior in the homogeneous liquid in more detail. Note that the pretransitional effect is negligible in the homogeneous phase and very strong in the two-phase region: The dashed line is for the CM law of rectilinear diameters, enabling the estimation of pretransitional anomaly strength: $\Delta_d = 25\%$.

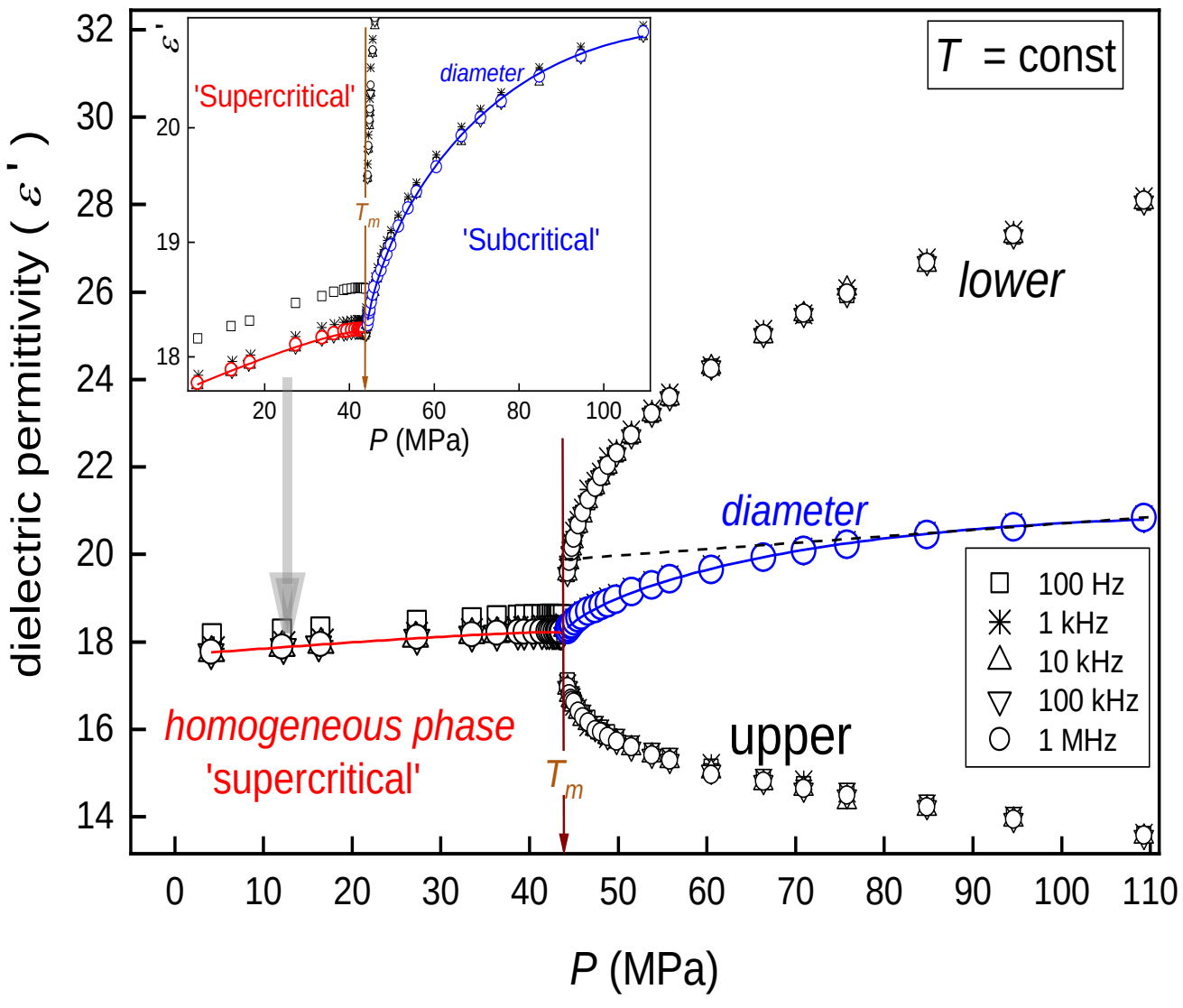

**Fig. 6** Pressure changes of dielectric permittivity $\varepsilon'$ (dielectric constant ($\varepsilon$) for $f$ >1 kHz) in the one and two-phase regions of nitrobenzene-dodecane mixture. For the coexistence curve, the location of the diameter is shown. The inset presents the behavior in the isotropic liquid phase in more detail. The dashed line shows the pressure counterpart of the CM law of rectilinear diameter, enabling the estimation of the pretransitional anomaly strength $\Delta_d$ = 9.7%; for its parallel in the homogeneous phase, one obtains: $\Delta_{iso.} = 0.7\%$ .

Figure 6 presents the isothermal pressure changes of dielectric permittivity in nitrobenzene dodecane critical mixtures. Studies were carried out for the isotherm located 5K above the critical consolute temperature under atmospheric pressure in the homogeneous liquid phase. There is clear evidence for the precritical effects described by the pressure counterpart of Eq. (15) [25, 41]:

$$\varepsilon'(P) = \varepsilon_C + a_P(P_c - P) + A_P(P_c - P)^{1-\alpha} \qquad (16)$$

where the exponent $\alpha = 0.12$.

Notable is the lack of correction-to-scaling terms and the possibility of describing changes of dielectric permittivity by the above relation even for as low frequency as 100 Hz, i.e., in the region dominated by the ionic mechanism in standard temperature tests: only the sift in the value of $\varepsilon_C$ appear. Such behavior agrees with earlier similar studies on compressing in the homogeneous phase of critical mixtures [41].

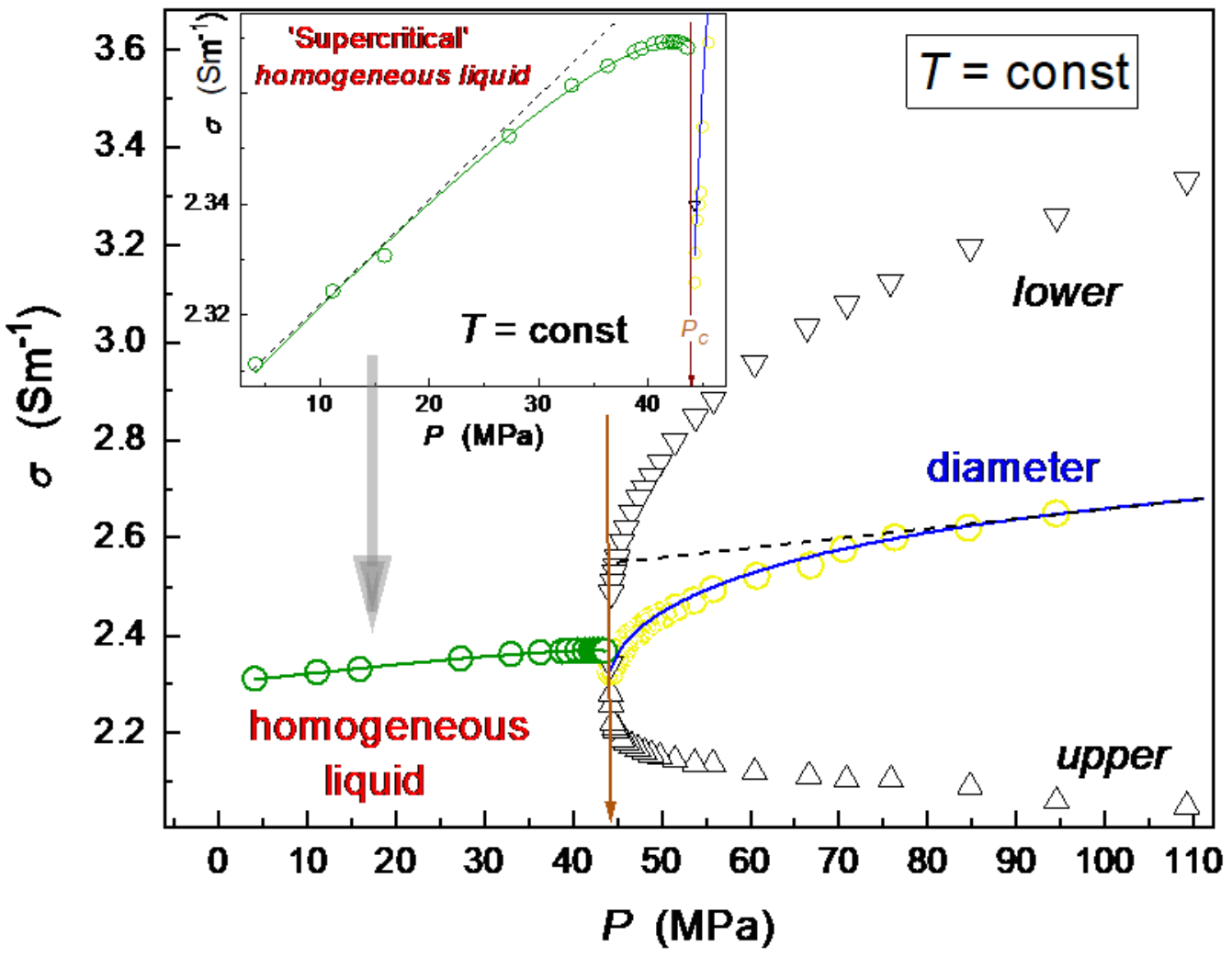

**Fig. 7** Isothermal, pressure-related changes of DC electric conductivity in nitrobenzene – dodecane critical mixture. The diameter of the coexistence curve is also shown. The instance is for the focused presentation of the behavior in the homogeneous liquid phase. Curves portraying precritical effects are related to Eqs. (16) and (17). The dashed line is for the CM law of rectilinear diameter (pressure parallel), enabling the introduction of the strength of the pretransitional anomaly: $\Delta_d = 10\%$, and the similar metric for the homogeneous phase: $\Delta_{iso.} = 0.7\%$.

The mentioned shift diminishes for $f > 1KHz$, and in this range $\varepsilon' = \varepsilon$ . The strong pretransitional anomaly of the diameter in the two-phase 'subcritical' domain is also notable. It is well portrayed by the pressure counterpart of Eq. (15a):

$$d_\varepsilon(P) = \frac{\varepsilon_{liquid}^{upper} + \varepsilon_{liquid}^{lower}}{2} = \varepsilon_C + B_d(P - P_C)^{2\beta} + A_d(P - P_C)^{1-\alpha} + a_d(P - P_C) \quad (17a)$$

$$d_\sigma(P) = \frac{\varepsilon_{liquid}^{upper} + \varepsilon_{liquid}^{lower}}{2} = \sigma_C + B_d^\sigma(P - P_C)^{2\beta} + A_d^\sigma(P - P_C)^{1-\alpha} + a_d^\sigma(P - P_C) \quad (17b)$$

where the exponent $\alpha = 0.12$.

Regarding DC electric conductivity, the pressure-related pretransitional anomaly is even stronger than the temperature path discussed above. Moreover, the explicit pretransitional anomaly appears also in the homogeneous ‘supercritical’ domain. It can be portrayed by the relation:

$$\sigma(P) = \sigma_C + a_P^{\sigma}(P_c - P) + A_P^{\sigma}(P_c - P)^{1-\alpha} \tag{16}$$

Results presented in Figs. (4 – 7) indicate that the application of binary mixtures of limited miscibility for extraction technologies can be very effective and ‘criticality supported’ for isothermal pressure paths. One can operate effectively in the two-phase domains, with the support of the ultrasound action. One should stress that the strong pretransitional anomaly of the diameter results from the impact of precritical fluctuations on ‘branches’ of the coexistence curve. It means that the beneficial impact of critical phenomena has to exist even for the ‘classic’ usage of the liquid-liquid extraction technology associated with the ‘dissolution’ but operating near the critical consolute point. The comparison of results presented in Figs. (4-7) also show that the pressure-related (isothermal) can be significantly more efficient than the classic, temperature-related operation under atmospheric pressure. First, pressure-related precritical anomalies are more robust than their temperature-related counterparts. Second, pressure can be changed ‘immediately’ and homogeneously, even for large-volume systems. It is not possible for temperature changes.

To conclude the discussion on the pressure-related critical LL extractions technology worth presenting are approximate pressure counterparts of the Kirkwood Eq. (7) and -Withney Eq. (8)

$$k(T,P), s(T,P) \approx \frac{P}{T}\frac{p_{\infty}\Delta_P}{R}\left(\frac{1}{\varepsilon(T)} - 1\right) + p_{\infty} = \frac{P}{T}K\left(\frac{1}{\varepsilon_C + a(P_c - P) + A(P_c - P)^{1-\alpha} + ..} - 1\right) + p_{\infty} \tag{17}$$

$$\frac{dm}{dt} = \frac{S\Delta_C}{d}\left(\sigma_c + a_{\sigma}(P_C - P) + A_{\sigma}(P_c - P)^{1-\alpha}\right) \tag{18}$$

where $T = const$ and in the given case, the prefactor $p_{\infty}$ is for the value under atmospheric pressure, for the tested temperature.

In the case of weakly-discontinuous phase transitions, singular temperatures, and pressure $(T^*, P^*)$ should appear instead of $(T_C, P_C)$ in Eqs. (7, 8) and (17, 18)

The simple extraction using a liquid carefully selected to the process target constitutes the simplest and probably the fundamental laboratory and industrial technology. It seems inherently free from any critical phenomena resulting from the Gibbs-Kohnstamm phase rule [42]. Pretransitional/precritical phenomena are coupled with continuous or weakly discontinuous phase transitions [11]. There is extensive experimental evidence, supported by fundamental theoretical reasoning, that there are no pretransitional effects in the liquid phase on approaching the melting/freezing discontinuous phase transition [10, 11]. Such transition occurs 'suddenly', when passing $T_{m,}/T_f$ and they are associated with 'jump-type' changes of physical properties: density, heat capacity, dielectric constant.

Figure 8 shows the unique exception from this rule, noted in the liquid phase of linseed oil. Linseed oil is natural vegetable oil with an enormous range of applications from food to cosmetics, pharmacy, and medicine. Its pro-health properties have been explored since ancient times [43]. Figure 8 presents the temperature change of dielectric constant in linseed oil, obtained from Golden Flax seeds. Presented dielectric constant changes show the strong pretransitional effect extending even 80 K above the melting temperature (!). The blue curve in Fig. 8 shows its fair portrayal by Eq. 8, the same as for the isotropic phase of rod-like liquid crystalline materials, but the extraordinary small value of the phase transition discontinuity $\Delta T^* = T_m - T^* \approx 0.6K$.

Notable that the pretransitional anomaly also appears in the solid phase. It can be associated with the dielectric response of liquid nano/microlayers between solid crystalline grains appearing just below the melting temperature. Its portrayal offers the relation [44]:

$$\varepsilon(T < T_C) = C + \frac{A}{T^{**} - T} \qquad (19)$$

where $C$ is constant and $T^{**}$ is the extrapolated temperature of a hypothetical continuous transition, denoted by 'star' in Fig. 8.

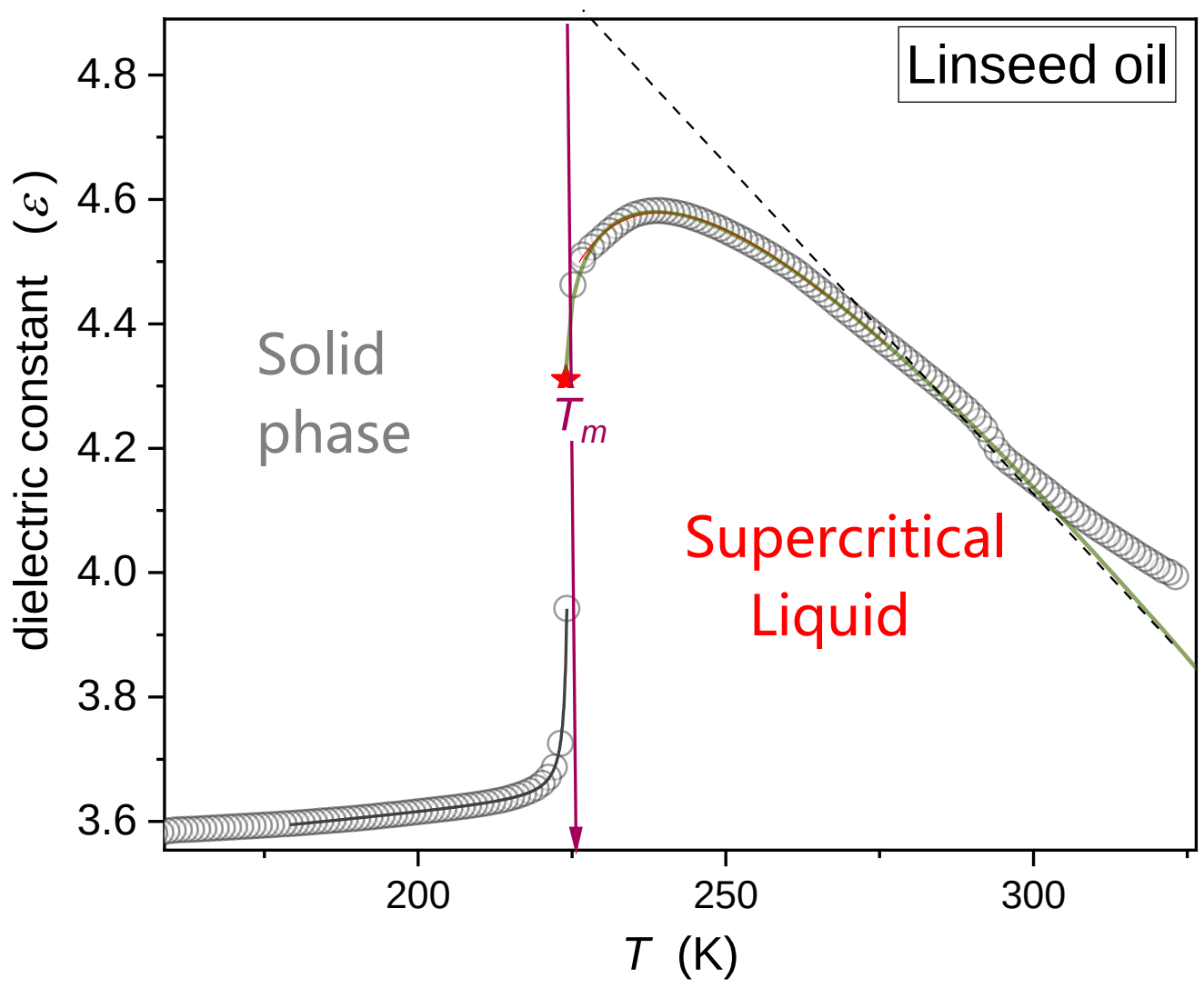

**Fig. 8** Changes of dielectric constant on cooling in linseed oil. Curves portraying the pretransitional behavior are related to Eq. (4), with the exponent $\alpha \approx 0.5$ for the liquid phase and Eq. (19) for the solid phase. The dashed line shows the linear extrapolation of the behavior remote from the singular temperature, enabling the estimation of the pretransitional anomaly strength $\Delta_{iso.} = 16\%$ .

Consequently, the question arises if superior pro-health features of linseed oil are not supported by extraordinary features of the supercriticality, recalled above?

The name 'supercritical' is here applied for any liquid/fluid system above a hypothetical 'critical point'. Its description should be governed by the critical-type parameterization of properties essential for classic supercritical extraction, such as dielectric constant.

So far, unique features of linseed oil have been explained by the beneficial impact of the unique composition of oly-unsaturated fatty acids (PUFA), phytoestrogenic lignans (seco-isolariciresinol diglucoside, SDG), and antioxidants such as phenolic acids and flavonoids [43]. The long-range and 'strong' pretransitional changes of dielectric constant can also suggest that linseed oil can constitute the base for the new generations of liquid extraction supported by supercriticality. Notably, linseed oil is a natural product without a hazardous impact on human health.

## 5. Conclusions

This report presents the results of studies that can constitute the fundamental base and inspiration for the new generation of SuperCritical Fluids (SCF) and Liquid-Liquid and Liquid extraction technologies, supported by critical phenomena. It is related to the shift $Supercritical\,(SCF, SuperCF\ \Rightarrow SubCF,$ due to the supplementary exogenic impact of pressure. The application of LL technology for binary mixtures of limited miscibility can be realized with the strong impact of critical fluctuations, i.e., linking beneficial features of classic SCF and LL concepts. In practice, this can be additionally supported by the impact of ultrasound or/and by isothermal processing under pressure. Inherent features of binary mixtures of limited miscibility made it possible to carry out the process even well below 100 MPa (1 kbar) and at near-room temperatures, i.e., at technologically convenient conditions. The discovery of far-range supercritical effect on linseed oil may open new perspectives for the 'critical support' of liquid-based extraction processing, in line with modern trends searching for 'natural & sustainable' process carriers. The report also sketches the fundamental base of discussed innovative possibilities of extraction technologies developments.

### Acknowledgments

Studies were carried out due to the National Centre for Science support (ref. UMO-2017/25/B/ ST3/02458, headed S. J. Rzoska. The paper is associated with the *International Seminar on Soft Matter & Food – Physico-Chemical Models & Socio-Economic Parallels, 1st Polish-Slovenian Edition*, Celestynów, Poland, 22–23 Nov., 2021; directors: Dr. hab. Aleksandra Drozd-Rzoska (Institute of High Pressure Physics Polish Academy of Sciences, Warsaw, Poland) and Prof. Samo Kralj (Univ. Maribor, Maribor, Slovenia).

**Conflict of interests:** the authors declare no conflict of interests.

**Authors' contributions:** the authors declare equal contributions.